\documentclass{conference}

\usepackage{graphicx}
\usepackage{amssymb}

\begin{document}

\begin{frontmatter}
\title{The theory of focusing of high energy ions by bent crystals of special shape}

\author{Gennady V. Kovalev}
\author{}
\address{School of Mathematics \\University of Minnesota, Minneapolis, MN 55455,USA}

\begin{abstract}
In this report we present the detail theory of the focusing in bent crystals using the statistical mechanics. The quantum mechanical effects for focusing and abberations are not considered. The beam structure in phase space are consedered and envelope near the focusing spot and intensity profile are derived.
\end{abstract}

\end{frontmatter}

\section{\label{sec:level1}Introduction}

\begin{figure}
\includegraphics{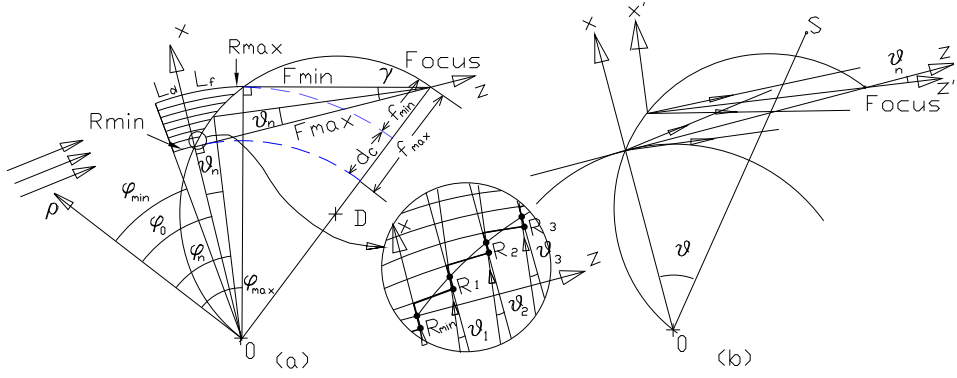}% Here is import EPS art
\caption{\label{fig:normal} Geometry of focusing crystal for high
energy particles (angles, thicknesses and curvatures are
exaggerated). (a) The radius-vectors $R_0, R_1,...R_N$ originated
from center $O$ designate the curvature of each plane. (b)The
$n$-th channel ($n=1,2,3...N$) is located between $n$-th and
$(n+1)$-th planes and has a 'local' Cartesian coordinate system
$(x',z')$ with offset $(x_n,z_n)$ and angle direction
$\vartheta_n=\varphi_n-\varphi_0$ relatively to 'base' Cartesian
coordinate system $(x,z)$ of the channel $0$.}
\end{figure}

When a high energy particle with a momentum $P$ are captured in a
channeling motion of bent crystal, the equivalent magnetic field
acting on the particle can be estimated as  $\sim P/(0.3R)$, where
$R$ is a curvature radius of the bent crystal~\cite{ts76,carr97}.
For the momentums $P\gtrsim 300 GeV/c$ and radius $R \thicksim 1m$
the equivalent magnetic field can reach the values
$\gtrsim\!1000$~tesla. This phenomena has been extensively
investigated for a beam deflection, splitting, extraction, and
spin precession measurements\cite{carr97,bck,tar98}. The first
experiment of the bent channeling\cite{ad79} confirmed that the
bent crystal could also be used for constructing a focusing
element with extremely short focal length and small focal spot.
Several crystal devices for focusing were
suggested\cite{carr80,carr87,sun87,sg91}, but the successful
experiments\cite{denis92,baran92} were done with the crystal
having a special shape drawn in Fig.~\ref{fig:normal}(a). This is
a regular bent crystal with the output face carved in such a way
that the tangent lines to the planar channels on that face are
converged to a focal point. With these conditions, the center
points of the channels on the output face of the crystal
constitute a cylindrical surface with diameter $D$, greater than
curvature radii of bending, $D \geq R_{max} \geq R \geq R_{min}$.
The estimations\cite{kov03} show that the size of smallest focal
spot is proportional to the square root of the crystal thickness
$d_c$. There is also an evaluation of the intensity at the focal
spot, and the crystal geometry is proposed where the focal
parameters can reach the extreme values.

In this article we present the detail theory of the focusing in
bent crystals using the statistical mechanics. The quantum
mechanical effects for focusing and aberration are not considered
here.

\section{\label{sec:level2}The model potential of focusing by bent crystal}

We consider the classical model of channeling~\cite{lindhard65}
with a continuous temperature dependent planar potential, when
ions are captured into channeling trajectories and directed along
the bent atomic channels. Similar considerations in bent crystals
without boundary conditions were done by many
authors~\cite{tar80,kudo81,ellison82,ellison87,tar98} and based on
the assumption that curvatures of crystalline structure are small
and bent plane locally looks like flat surface. In local
coordinate system the model of continuous straight planar
potential~\cite{lindhard65} can be apply and describe the most
important features of bent channeling~\cite{tar98}. If each plane
has a constant curvature, it is convenient to use a polar
coordinate system $(\rho,\varphi)$ with a center located at the
point $O$ (see Fig.~\ref{fig:normal}(a))and direction
$(\rho,\varphi=0)$ tangent to the cylindrical surface. In such
coordinate system the crystal potential for the bent channeling in
Moliere's approximation \cite{molier47,app67,gem74} with a
different length of the channels can be described as following

\begin{eqnarray}
U_{b-c}(\rho,\varphi)=\sum_{n=0}^{N} U_{n}(\rho-R_n,\varphi),
\label{pot_c}
\end{eqnarray}
where $R_n=R_{min}+n*d_p$ is the radius of curvature of $n$-th
bent plane ($R_0=R_{min}, R_N=R_{max}$), $d_p$ is interplanar
distance, $U_{n}(\rho-R_n,\varphi)$ is temperature-dependent
continuum potential of this plane derived from the Moliere
potential~\cite{poizat71,barrett71} for straight plane:
\begin{eqnarray}
U_{n}(\xi,\varphi)=\left\{
\begin{array}{ll}
\overline{U}\sum_{i=1}^{3}\frac{\alpha_i}{\beta_i}\exp(\beta_i^2\frac{u^2}{2a^2})\\
(\exp(\beta_i\frac{\xi}{a})erfc(\beta_i\frac{u}{\sqrt{2}a}+\frac{\xi}{\sqrt{2}u})\\
+\exp(-\beta_i\frac{\xi}{a})erfc(\beta_i\frac{u}{\sqrt{2}a}-\frac{\xi}{\sqrt{2}u})),& \varphi_{min} < \varphi <\varphi_n,\quad \label{pot_M}\\
0, &  \varphi<\varphi_{min},\quad \varphi>\varphi_n.\\
\end{array} \right.
\end{eqnarray}

Here $a$ and $u$ are Thomas-Fermi screen radius and
root-mean-square temperature displacement of crystal atoms (in Si
at 293K $u=0.075 \AA)$, $\alpha_i=\{.1,.55,.35\}$, $\beta_i=\{6.0;1.2;.3\}$,
$\overline{U}=\pi Z e^2 a n_p$ ($n_p$ is a density of atoms in the
plane), $erfc(x)$ is the complementary error function. The angles
$\varphi_{min}$, $\varphi_{n}$  denote the enter and exit surfaces
of the $n$th bent plane. Thus, the end cylindrical surface of the
crystal lies in the range $\varphi_{0} <\varphi <\varphi_{max}$.
Others notations can be found in Fig.~\ref{fig:normal}. Although
Moliere's approximation ~(\ref{pot_M}) for isolated plane has a
simple analytical form, it can not directly be implemented for
calculation of the channeling trajectory, because the channel
potential is superposed from several neighboring planes. However,
if we consider an infinite crystal at fixed temperature, this
superposition can be fitted by a polynomial of suitable order. The
estimations can be easily received by an harmonic
approximation~\cite{tar98}, which coincides with Moliere's
approximation on the crystalline planes and on centers of the
channels. By involving 10 crystalline planes located far away from
boundaries ($K>>1$), the parabolic potential can be fitted to the
Moliere's approximation with the following parameters:
\begin{eqnarray}
U_{max}=\sum_{n=K-4}^{K+5} U_{n}((K-n)*d_p,\varphi),\nonumber\\
U_{min}=\sum_{n=K-4}^{K+5}
U_{n}((K-n)*d_p-d_p/2,\varphi).\label{fit}
\end{eqnarray}

\begin{figure}
\includegraphics{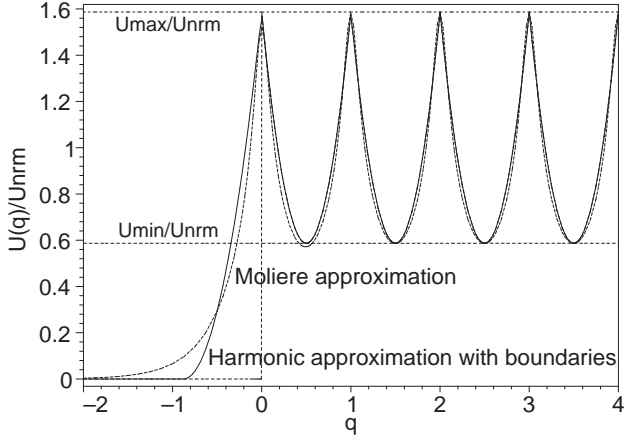}% Here is import EPS art
\caption{\label{fig:Si110Pot} The crystal potential for the range
$\varphi_{min} < \varphi <\varphi_0$ in harmonic approximation
with parameters $U_{max}=35.9 eV,\quad U_{min}=13.3 eV$, and
Moliere potential for Si (110) at T=300K. Notations:
$q=(\rho-R_{min})/d_p, U_{nrm}=U_{max}-U_{min}$.}
\end{figure}

One can extend this model to the crystalline potential near the
boundaries ignoring the fact that potential of the first several
subsurface channels becomes slightly asymmetric (see
Fig.~\ref{fig:Si110Pot}). Outside the crystal, the potential can
also be fitted by half part of parabola, for example, with a
condition that on the boundary plane the potential is continuous
and the parabola intersects the Moliere's potential at the value
$U_{min}/2$. Because other planes give a small contribution to the
surface potential, the distance of this intersection from boundary
plane is very close to $d_p/2$. The resulting system of two
equations,
\begin{eqnarray}
U_b*(\rho_b/d_p+1/2)^2=U_{min}/2,\nonumber\\
U_b*(\rho_b/d_p)^2=U_{max},\label{boundary}
\end{eqnarray}
leads to the boundary semiparabolic potential with parameters:
\begin{eqnarray}
U_b=\frac {(2U_{max}-U_{min})^2}{U_{max}+U_{min}/2+\sqrt{2U_{max}U_{min}}},\nonumber\\
\rho_b=-d_p\frac {U_{max}+\sqrt {U_{max} U_{min}/2}}{2\,{\it
U_{max}}-\,{\it U_{min}}}. \label{boundary semiparabolic potential}
\end{eqnarray}

Finally, the parabolic model of Moliere's  potential
~(\ref{pot_M}) of the bent crystal with boundaries can be written
(see Fig.~\ref{fig:Si110Pot})
\begin{eqnarray}
U_{0}(\rho,\varphi)=\left\{
\begin{array}{ll}
U_{b}\left(\frac{\rho-R_0-\rho_b}{d_p}\right)^2,& \rho_{b}+R_{0}
< \rho < R_{0},\varphi_{min} < \varphi <\varphi_0,\quad  \\
0, &   otherwise, \label{pot_bo}\\
\end{array} \right.
\end{eqnarray}

- for boundary and

\begin{eqnarray}
U_{n}(\rho,\varphi)=\left\{
\begin{array}{ll}
U_{nrm}\left(\frac{2(\rho-R_n-d_p/2)}{d_p}\right)^2+U_{min},&
R_{n} < \rho <R_{n+1},\\
& \varphi_{min} < \varphi <\varphi_n,\quad  \\
U_{b}\left(\frac{\rho-R_{n+1}-\rho_b}{d_p}\right)^2,&
\rho_{b}+R_{n+1}
< \rho < R_{n+1},\\
& \varphi_{n} < \varphi <\varphi_{n+1},\quad  \\
0, &  otherwise,  \label{pot_h}\\
\end{array} \right.
\end{eqnarray}

- for channels $ 0\leq n \leq N-1$. The boundary potential
~(\ref{pot_bo}) does not play any role for bulk channeling, but
may play an essential role for an abberation of the focusing, if
the length of this potential along the n-th microbeam, $R_n
(\varphi_{n+1}-\varphi_{n})$, greater then the length of
oscillation of particles inside the channel, $\lambda_c \approxeq
d_p/\Psi_p$ ($\Psi_p$ is the Lindhard critical angle for planar
channeling). This condition may happen for extreme focusing with
shortest focal length ~\cite{kov03} (see also below) and a similar
potential term at the end of each channel must be kept for
corrections of phase curves (see second line of ~(\ref{pot_h})).
If this condition does not hold and distortion caused by
boundaries is negligible, and the channel potential
~(\ref{pot_bo},\ref{pot_h}) becomes:
\begin{equation}
U_{b-c}(\rho,\varphi) = \sum_{n=0}^{N-1} U_{n}(\rho,\varphi), \label{pot_c2}\\
\end{equation}
where
\begin{eqnarray}
U_{n}(\rho,\varphi)=\left\{
\begin{array}{ll}
U_{nrm}\left(\frac{2(\rho-R_n-d_p/2)}{d_p}\right)^2+U_{min},&
R_{n} < \rho <R_{n+1}, \\
&  \varphi_{min} < \varphi <\varphi_n,\quad  \\
0, &  otherwise.  \label{pot_p}\\
\end{array} \right.
\end{eqnarray}

\section{\label{sec:level3}Dynamic of particles inside the bent crystal}

Consider the motion of a relativistic spinless particles in
potential ~(\ref{pot_c}) or ~(\ref{pot_c2}). The Hamiltonian of
the particle is defined by
\begin{equation}
H(Q,P)=c\sqrt{P_i P^i+m^2c^2}+U_{b-c}(\rho,\varphi),
\label{hamiltonian}
\end{equation}

where $Q=\{\rho,\varphi\}$ and $P=\{p_{\rho},p_{\varphi}
\}=\{m\dot{\rho}\gamma,m\rho^2\dot{\varphi}\gamma \}$ are
canonical coordinates and conjugate momenta of particle, $\gamma
=(1-(\dot{\rho}^2+\rho^2\dot{\varphi}^2)/c^2)^{-1/2}$, $P_i
P^i=p_{\rho}^2+p_{\varphi}^2/\rho^2$. The Hamiltonian $H(Q,P)$ is
a piecewise function of $\varphi$ and $\rho$ and, most important,
does not depend on $\varphi$ inside and outside of each individual
channel. Hence, the angular momentum of particle
$p_{\varphi}=m\rho^2\dot{\varphi}\gamma $ (see
Fig.~\ref{fig:normal}) is conserved inside and outside the crystal
and may change only when particle crosses the boundaries of the
crystal. The phase curves of the particles are described by system
of Hamiltonian's equations ( $\dot{P}=-\partial H(Q,P)/
\partial Q,\quad \dot{Q}=\partial H(Q,P)/
\partial P$):
\begin{eqnarray}
\dot{p_{\rho}}=\frac{cp_{\varphi}^2}{\rho^3\sqrt{P_i
P^i+m^2c^2}}-\partial
U_{b-c}(\rho,\varphi)/\partial \rho, \quad \dot{p_{\varphi}}=0, \nonumber\\
\dot{\rho}=\frac{cp_{\rho}}{\sqrt{P_i P^i+m^2c^2}},\quad
\dot{\varphi}=\frac{cp_{\varphi}}{\rho^2\sqrt{P_i
P^i+m^2c^2}}.\label{hamilton1}
\end{eqnarray}
Using the fact, that last two equations of ~(\ref{hamilton1}) can
only be resolved if

\begin{equation}
\frac{c}{\sqrt{P_i P^i+m^2c^2}}=\frac{1}{m \gamma},
\end{equation}

the system ~(\ref{hamilton1}) can be written in the form:

\begin{eqnarray}
\dot{p_{\rho}}=\frac{p_{\varphi}^2}{m\gamma\rho^3}-\partial
U_{b-c}(\rho,\varphi)/\partial \rho, \quad p_{\varphi}=const, \nonumber\\
\dot{\rho}=\frac{p_{\rho}}{m\gamma},\quad
\dot{\varphi}=\frac{p_{\varphi}}{\rho^2m\gamma}. \label{hamilton2}
\end{eqnarray}
The Hamiltonian (\ref{hamiltonian}) of the system
(\ref{hamilton2}) has a cylindrical symmetry, therefore the
angular momentum of the particle $p_{\varphi}$ is conserved. The
tangent momentum of particle $p_{\varphi}/\rho$ along the curved
channel is not exactly conserved, but taking into account that the
channeling particle can not move further than $d_p/2$ from the
center line of the channel and $d_p/R_{n} \ll 1$, the tangent
momentum of particles
$p_t=p_{\varphi}/\rho=m\rho\dot{\varphi}\gamma$ and tangent
velocity $v_t=\rho\dot{\varphi}=p_t/(m\gamma)$ are conserved with
high accuracy. We also make the following

\begin{figure}
\includegraphics{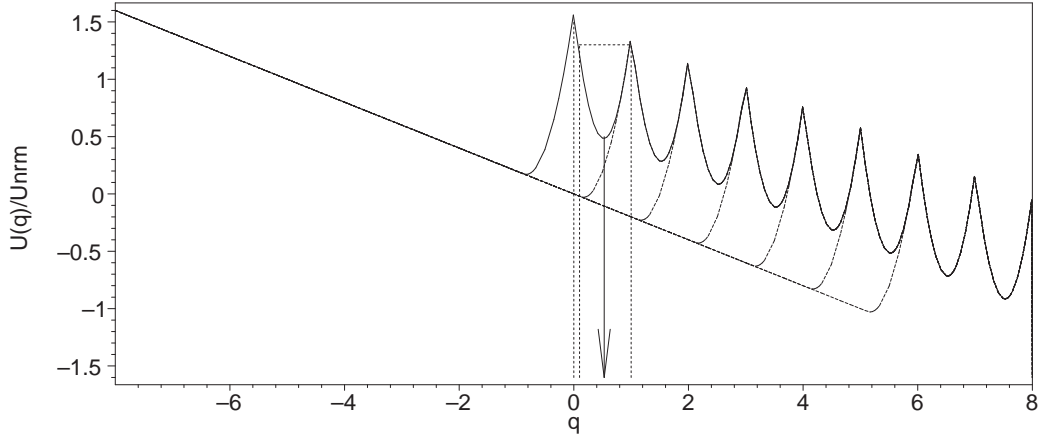}% Here is import EPS art
\caption{\label{fig:Si110PotEff} The effective crystal potential
for bounded crystal Si (110) at T=300K in harmonic approximation,
Eq.~(\ref{pot_bo},\ref{pot_h}),  for radial sections:
$\varphi_{min} < \varphi <\varphi_0$ - solid line; $\varphi_{0} <
\varphi <\varphi_1$ and $\varphi_{1} < \varphi <\varphi_2$ - dash
lines. In the bent crystal,  in addition to the normal channels,
particle may also have a surface channel in which it moves
sequentially reflecting from a surface plane. Note that this
surface channel is much wider than the crystal channel and might play an
important role for focusing in extreme conditions ~\cite{kov03}.}
\end{figure}

approximations:
\begin{eqnarray}
p_{\rho} \ll (p_t, mc), \nonumber\\
\frac{p_{\varphi}^2}{m\gamma\rho^3} \simeq \frac{p_{\varphi}^2}{m\gamma R_{min}^3},\nonumber\\
U_{b-c}(\rho,\varphi) \ll c\sqrt{P_i
P^i+m^2c^2},\label{assump}\\
\nonumber
\end{eqnarray}
which mean that the incident particles are directed at the small
angle to the crystalline planes, the centrifugal force
$p_{\varphi}^2/(m \gamma \rho^3)=p_{t}v_{t}/\rho$ for all crystal
channels is practically the same (the crystal thickness
$d_c=Nd_p\ll R_{min}$), and gamma factor inside the crystal is
changed by a small amount:
\begin{equation}
\gamma \simeq \gamma_o = \frac{\sqrt{p_{\rho o}^2+p_{\varphi
o}^2/\rho_o^2+m^2c^2}}{m c}, \label{gamma}
\end{equation}
From ~(\ref{hamilton1}), (\ref{assump}) and ~(\ref{gamma}) the well known equations ~\cite{ts76} of motion of a particle in bent crystal can easily be received :

\begin{eqnarray}
m\gamma_o \ddot{\rho}=\frac{p_{\varphi}^2}{m\gamma_o R_{min}^3}-\partial U_{b-c}(\rho,\varphi)/\partial \rho, \quad p_{\varphi}=const,\nonumber\\
\dot{\rho}=\frac{p_{\rho}}{m\gamma_o},\quad
\dot{\varphi}=\frac{p_{\varphi}}{R_{min}^2 m\gamma_o}.
\label{hamilton4}
\end{eqnarray}

Note that parameter $\varphi$ in
Eq.~(\ref{hamilton1},\ref{hamilton2},\ref{hamilton4})
means only the piecewise character of potential~(\ref{pot_c},
\ref{pot_c2}), which has discontinuity at the front and end
surfaces of the each channel. The normal derivative of the
potential $\partial U_{b-c}(\rho,\varphi)/\partial \varphi$ on
these surfaces does not exist. Of course, the real atomic
potential is smooth function on the surface. Its derivative
rendering the normal force causes an impulse in normal linear
momentum when the particles cross these surfaces. The linear
momentum tangent to the front and end surfaces of the channels
does not change as particle passes the boundary of the crystal,
but angular momentum may slightly change due to force acting
perpendicular to these surfaces. Thus, the boundary conditions for
Eq.(~\ref{hamilton1},\ref{hamilton2}-\ref{hamilton4}) on front
surface of the crystal, $\varphi = \varphi_{min}$, can be
presented in the form
\begin{equation}
\rho_o=\rho_i, \quad  p_{\rho o}=p_{\rho i}, \quad p_{\varphi
o}=p_{\varphi i}+\Delta_{p}, \label{boundary_front}
\end{equation}
where   $(\rho_o,p_{\rho o},p_{\varphi o})$, $(\rho_i, p_{\rho i},
p_{\varphi i})$ denote the radial coordinate,  linear tangent and
angular momenta on the particle for
$\varphi\rightarrow\varphi_{min}^-$ and
$\varphi\rightarrow\varphi_{min}^+$ respectively. Using the law of
conservation of the energy
\begin{equation}
c\sqrt{p_{\rho o}^2+p_{\varphi
o}^2/{\rho_o}^2+m^2c^2}=c\sqrt{p_{\rho i}^2+p_{\varphi
i}^2/{\rho_i}^2+m^2c^2}+U_{b-c}(\rho_i,\varphi_{min}),\label{consE}
\end{equation}
we can receive a change in angular momentum $p_{\varphi}$
in~(\ref{boundary_front}),
\begin{equation}
\Delta_{p}\cong \frac{\rho^2 \sqrt{p_{\rho o}^2+p_{\varphi
o}^2/\rho_o^2+m^2c^2}}{c p_{\varphi
o}}U_{b-c}(\rho_i,\varphi_{min}),\label{delta}
\end{equation}
on the boundary of the crystal. The same boundary conditions
~(\ref{boundary_front}) take place when the particle passes the end surface
of the crystal. The assumption ~(\ref{assump}) allows also to
split the energy of relativistic particles into kinetic energy of
particle along the bent channel $E_{||}$ and transverse energy
$E_{\perp}$ ~\cite{pathak76,kaplin78,tar80,tar81,ellison87}:
\begin{eqnarray}
E \cong E_{||} +E_{\perp}, \quad \quad \nonumber\\
E_{\perp}=\frac{p_{\rho}^2}{2m \gamma_o} +
U_{b-c}(\rho,\varphi) -\frac{p_\varphi^2 (\rho-R_{min})}{m \gamma_0 R_{min}^3 },\ 
E_{||}=c\sqrt{p_{\varphi}^2/R_{min}^2+m^2c^2}.
\label{par_tran}
\end{eqnarray}
The transverse energy $E_{\perp}$ is constructed of  three terms: the kinetic transverse energy, the static potential energy and centrifugal potential energy.
The last two terms may be combined in one - the effective transverse potential
in a bent crystal (see Fig.~\ref{fig:Si110PotEff})):
\begin{equation}
U_{eff}(\rho,\varphi)=U_{b-c}(\rho,\varphi)-\frac{p_\varphi^2 (\rho-R_{min})}{m \gamma_0 R_{min}^3 }.
 \label{pot_eff}
\end{equation}

Since the energy along the bent channel $E_{||}$ is conserved, the
transverse energy $E_{\perp}$ is also conserved. Equation
(\ref{hamilton4}) with the potential (\ref{pot_c2},\ref{pot_p})
and boundary conditions (\ref{boundary_front},\ref{delta}) can be easily
resolved,
\begin{eqnarray}
\rho =(\rho_o-\rho_{c,n}) \cos(\omega t)+ \frac{p_{\rho o}}{\omega m \gamma_o}\sin(\omega t)+\rho_{c,n}, \nonumber\\
p_{\rho}=p_{\rho o}\cos(\omega t)-(\rho_o-\rho_{c,n}) \omega m
\gamma_o \sin(\omega t),\nonumber\\
\varphi=\Omega t +\varphi_{min},\nonumber\\
p_{\varphi}=const,
\label{sol}
\end{eqnarray}

where the  frequency of angular rotation, frequency of transverse oscillation   and the radius of the central line of motion for $n$-channel are

\begin{eqnarray}
\Omega = \frac{p_{\varphi}}{m \gamma_o R_{min}^2},\nonumber\\ 
\omega=\sqrt{\frac{8U_0}{m \gamma_o d_p^2}}, \nonumber\\
\rho_{c,n}=R_n+\frac{d_p}{2}+\frac{p_{\varphi}^2 }{\omega^2 m^2
\gamma_o^2 R_{min}^3}.
\label{fr2}
\end{eqnarray}

Thus,  the motion of all particles in phase
space can be described by Eqs. (\ref{sol}) with parameters (\ref{fr2}).
The maximum amplitude of oscillation and transverse momentum inside the bent crystal are (see Eq.(\ref{sol}) and Fig.\ref{fig:Si110PotEff})
\begin{eqnarray}
|\rho_{max}-\rho_{c,n}|=\frac{d_p}{2}-\frac{p_{\varphi}^2 }{\omega^2 m^2
\gamma_o^2 R_{min}^3},\nonumber\\
p_{\rho,max}=(\frac{d_p}{2}-\frac{p_{\varphi}^2 }{\omega^2 m^2
\gamma_o^2 R_{min}^3})\omega m
\gamma_o.
\label{MaxAmplitude}
\end{eqnarray}

To facilitate further calculations we introduce the normalized  dimensionless
local coordinates $x,v_x$ for each channel:

\begin{eqnarray}
x = \frac{\rho - \rho_{c,n}}{d_p/2},\  
v_x = \frac{p_{\rho}}{\omega m \gamma_o d_p/2}, 
\label{CanonicalCoords}
\end{eqnarray}

which satisfy the equation of conservation transverse energy (\ref{par_tran}) 

\begin{eqnarray}
x^2 + v_x^2 = r_{ch}^2. 
\end{eqnarray}

The parameter $r_{ch}$,

\begin{eqnarray}
r_{ch} =1-\frac{2 p_{\varphi}^2 }{\omega^2 m^2 \gamma_o^2 R_{min}^3 d_p}=1-\frac{\epsilon}{\epsilon_c}, \nonumber\\
\epsilon = \frac{p_{\varphi}^2}{m \gamma_o R_{min}^3},  \epsilon_c = \frac{4 U_0}{d_p}, 
\label{RadiusPhase space}
\end{eqnarray}

can be called the dimensionless radius of the microbeam in the phase space, and
$\epsilon$, $\epsilon_c$ are an effective electric field and critical electric field produced by bent crystal ~\cite{tar98,bck}. 
\section{\label{sec:level4}Statistical model of microbeam focusing}

The classical approach to the kinetics of beams assumes the Boltzmann equation for
particle distribution function $f = f(\overrightarrow{r},
\overrightarrow{p})$ in a steady state situation:
\begin{equation}
\overrightarrow{v}\frac{\partial f}{\partial \overrightarrow{r}}+
e \overrightarrow{E}\frac{\partial f}{\partial
\overrightarrow{p}}=St(f),
\end{equation}
where $e  \overrightarrow{E}$ is electric field acting upon the
ion with charge $e $ inside the crystal, $St(f)$ is a collision
integral, which plays an important role in establishing the
equilibrium states and dechanneling inside the crystal. Outside the crystal $St(f)$ and $e  \overrightarrow{E}$ vanish and the Boltzmann equation reduces to particular simple form 
\begin{equation}
\overrightarrow{v}\frac{\partial f}{\partial \overrightarrow{r}}=0.
\label{Free_Boltzman}\\
\end{equation}
Consider 2-D geometry for plane channeling (Fig.~(\ref{fig:normal}a)). The particle beam has an initial distribution $f_{in}(x, z_{in}(x),v_{x},v_{z})$ on the entrance of the crystal
surface $z = z_{in}(x)$ and some distribution $f_{out}(x,z_{out}(x),v_{x},v_{z})$ formed by channeling process on the end face of the crystal $z = z_{out}(x)$. We use Cartesian coordinate system instead of the cylindrical coordinates used in Sec.\ref{sec:level2},\ref{sec:level3} and assume that all coordinates and their conjugate momentums or velocities are dimensionless. Behind the crystal the Eq.~(\ref{Free_Boltzman}) becomes 
\begin{equation}
v_x\frac{\partial f(x,z,v_{x},v_{z})}{\partial x}+ v_z\frac{\partial f(x,z,v_{x},v_{z})}{\partial z}=0
\label{Free_Boltzman2D}\\
\end{equation}
with general solution 

\begin{equation}
f(x,z,v_{x},v_{z})=g(x- z*v_x/v_z,v_{x},v_{z}), 
\label{General_Solution2D}\\
\end{equation}

where $g$ is arbitrary function of three variables. The boundary condition on the end face of the crystal $z=z_{out}(x)$ gives unique solution for beam distribution ($v_z=const$):  

\begin{equation}
g(x- z_{out}(x)*v_x/v_z,v_{x},v_{z})=f_{out}(x,z_{out}(x),v_{x},v_{z}).
\label{Boundary_Boltzman2D}\\
\end{equation}

\begin{figure}
	\centering
		\includegraphics{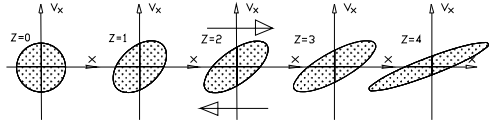}
	\caption{Evolution of distribution function for microbeam from single channel. The trajectories of particle in 2D are parallel to the $x$ axis. The phase profile along trajectories of particles  moves with a constant velocity ($v_x =const$).}
	\label{fig:PHASE_CH}
\end{figure}

As a simple model, consider single channel, say channel with output angle $\varphi_0$ (see Fig.\ref{fig:normal}a). Output particle distribution on the end face of the channel ($z_{out} = 0$) is 

\begin{eqnarray}
f_{out,0}(x,0,v_{x},v_{z})=h(v_z)\left\{
\begin{array}{ll}
\left. \frac{1}{\pi r_{ch}^2} \right. , & v_{x}^2+x^2 
< r_{ch}^2,\quad  \\
0, &  v_{x}^2+x^2 \geq r_{ch}^2, \\
\end{array} \right.
\label{Boundary_OneChan}
\end{eqnarray}

where $h(v_z)$ is some distribution in space of longitudinal component of the velocity, $r_{ch}$ is the radius of particle distribution in the phase space  which is less than $d_p/2$ radius of crystal channel. The  distribution (\ref{Boundary_OneChan}) is normalized 

\begin{equation}
\int_{-\frac{2}{d_p}}^{\frac{2}{d_p}} \int_{-\infty}^{\infty}\int_{-\infty}^{\infty}
f_{out,0}(x,0,v_{x},v_{z}) d x d v_{x} d v_{z}= 1, \label{NormalizationDistibutionOneCh}
\end{equation}

viz., one particle per one channel. By using boundary conditions (\ref{Boundary_Boltzman2D}) and (\ref{Boundary_OneChan}) it is easy to show that the distribution function of this channel in phase space behind the crystal will be

\begin{eqnarray}
f_{out,0}(x,z,v_{x},v_{z})=h(v_z)\left\{
\begin{array}{ll}
\left. \frac{1}{\pi r_{ch}^2} \right. , & v_{x}^2+(x-z\frac{v_x}{v_z})^2 
< r_{ch}^2,\quad  \\
0, &  v_{x}^2+(x-z\frac{v_x}{v_z})^2 \geq r_{ch}^2, \\
\end{array} \right.
\label{Distribution_FistChan}
\end{eqnarray}

and the evolution of this distribution function along the axis $z$ is shown in Fig.\ref{fig:PHASE_CH}. In phase space, the cross section of distribution is deformed, but the area of microbeam is conserved as well as the phase density along trajectories in the agreement with Liouville's theorem. 

The same treatment can easily be extended to each channel if we take proper Cartesian coordinate system $(\acute{x},\acute{z})$ with origin at the center of this channel on the end face of the crystal and with $OX$ axis directed to the focus (see Fig.\ref{fig:normal}b). The coordinate system for $n$ channel is connected to Cartesian coordinates $(x,z)$ of the 1st channel by following relations

\begin{eqnarray}
x -x_n = \acute{x}_n cos(\vartheta_n)-\acute{z}_n sin(\vartheta_n), \nonumber\\
z -z_n = \acute{x}_n sin(\vartheta_n)+\acute{z}_n cos(\vartheta_n),
\label{transform}
\end{eqnarray}

where the origins of $(\acute{x},\acute{z})$ coordinate systems are

\begin{eqnarray}
x_n = (R_{min}+ n d_p + d_p/2)cos(\vartheta_n)-R_{min}), \nonumber\\
z_n = (R_{min}+ n d_p + d_p/2)sin(\vartheta_n),
\label{origins}
\end{eqnarray}

and the angles $\vartheta_n =\varphi_n-\varphi_0$ defined by (see
Fig.\ref{fig:normal}a)

\begin{equation}
sin(\vartheta_n)=\frac{(R_{min}+ n d_p)\sqrt{D^2-R^2_{min}}-R_{min}\sqrt{D^2-(R_{min}+n d_p)^2}}{D^2}. 
\label{ThetaAngles}
\end{equation}

Among others important parameters of the focusing (Fig.\ref{fig:normal}), there are simple formulas for the maximum and minimum focal distances

\begin{eqnarray}
F_{max} = \sqrt{D^2-R_{min}^2}, \nonumber\\
F_{min} = \sqrt{D^2-(R_{min}+d_c)^2},\nonumber\\
d_c = N d_p.
\label{focal_distances}
\end{eqnarray}

\begin{figure}
	\centering
		\includegraphics{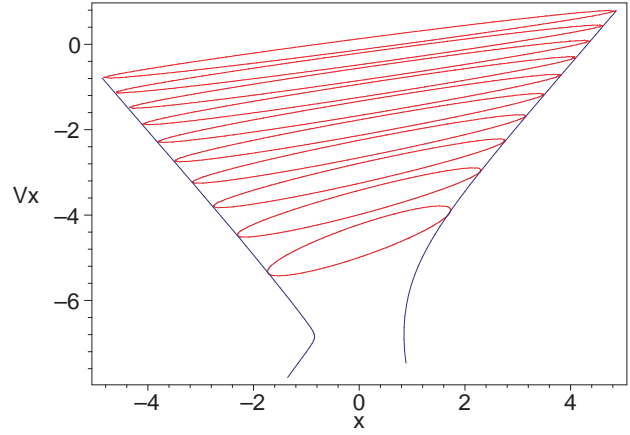}
	\caption{The phase space of microbeams at focal spot when $F_{min} << F_{max}$. Only ten channels are taken for illustrative purposes. In fact, about $10^7$ crystal (110) channels would typically be involve in forming a focal spot for the Si crystal with thickness $d_c = 2 mm$. }
	\label{fig:FocalSpot}
\end{figure}

The distribution function for n-th channel in phase space behind the crystal is

\begin{eqnarray}
f_{out,n}(\acute{x}_n,\acute{z}_n,v_{\acute{x}_n},v_{\acute{z}_n})=h(v_{\acute{z}_n})\left\{
\begin{array}{ll}
\left. \frac{1}{\pi r_{ch}^2} \right. , & v^2_{\acute{x}_n}+(\acute{x}_n - \acute{z}_n \frac{v_{\acute{x}_n}}{v_{\acute{z}_n}})^2 
< r_{ch}^2,\quad  \\
0, & v^2_{\acute{x}_n}+(\acute{x}_n - \acute{z}_n \frac{v_{\acute{x}_n}}{v_{\acute{z}_n}})^2 
\geq r_{ch}^2, \\
\end{array} \right.
\label{Distribution_n_Chan}
\end{eqnarray}

The equations (\ref{transform}) represent a reversible 
transformation, i.e. that they also define the $(\acute{x}_n,\acute{z}_n)$ as functions of the $(x,z)$, or, in other words, that they are soluble with respect 
to the $(\acute{x}_n,\acute{z}_n)$

\begin{eqnarray}
\acute{x}_n = (x -x_n) cos(\vartheta_n)+(z -z_n)sin(\vartheta_n), \nonumber\\
\acute{z}_n = -(x -x_n) sin(\vartheta_n)+(z -z_n) cos(\vartheta_n).
\label{transform_reverse}
\end{eqnarray}
 
For the distribution function (\ref{Distribution_FistChan}) we have also to calculate $v_{\acute{x}_n}$ and $v_{\acute{z}_n}$ which can be done by differentiating (\ref{transform_reverse}) with respect to $t$:

\begin{eqnarray}
v_{\acute{x}_n} = v_x cos(\vartheta_n)+ v_z sin(\vartheta_n), \nonumber\\
v_{\acute{z}_n} = -v_x sin(\vartheta_n)+ v_z cos(\vartheta_n).
\label{transform_velocities}
\end{eqnarray}

Note that the distribution in longitudinal space of the velocities is invariant under the changes of (\ref{transform_velocities}), i.e. $h(v_z)=h(v_{\acute{z}_n})$.
Substituting in the equation (\ref{Distribution_n_Chan}) the values of the $(\acute{x}_n,\acute{z}_n,v_{\acute{x}_n},v_{\acute{z}_n})$ given by formulas (\ref{transform_reverse}), (\ref{transform_velocities}), we get the equation for distribution functions for all channels in coordinates $(x,z,v_x,v_z)$

 \begin{eqnarray}
f_{out,n}(x,z,v_x,v_z)=h(v_z)\left\{
\begin{array}{ll}
\left. \frac{1}{\pi r_{ch}^2}\right. , &(v_x cos(\vartheta_n)+ v_z sin(\vartheta_n))^2 +\\
&\Big((x -x_n) cos(\vartheta_n)+(z -z_n)sin(\vartheta_n) - \\
&((x -x_n) sin(\vartheta_n)-(z -z_n) cos(\vartheta_n)) \\
&\frac{v_x cos(\vartheta_n)+ v_z sin(\vartheta_n)}{v_x sin(\vartheta_n)- v_z cos(\vartheta_n)}\Big)^2 
< r_{ch}^2,\quad  \\
\\
0, & (v_x cos(\vartheta_n)+ v_z sin(\vartheta_n))^2 +\\
&\Big((x -x_n) cos(\vartheta_n)+(z -z_n)sin(\vartheta_n) - \\
& ((x -x_n) sin(\vartheta_n)-(z -z_n) cos(\vartheta_n)) \\
&\frac{v_x cos(\vartheta_n)+ v_z sin(\vartheta_n)}{v_x sin(\vartheta_n)- v_z cos(\vartheta_n)}\Big)^2 
\geq r_{ch}^2,
\end{array} \right.
\label{Distribution_All_Chan}
\end{eqnarray} 

The final distribution function for the beam behind the crystal is 

\begin{equation}
f_{out}(x,z,v_x,v_z)=\sum_{n=0}^{N-1} f_{out,n}(x,z,v_x,v_z), 
\label{TotalDistribution}
\end{equation}

where $N = d_c/d_p$ - number of plane channels of crystal. If we substitute $z=F_{max}$ in (\ref{TotalDistribution}), we can get the distribution function in a focal plane of the phase space. The main features of this distribution can be seen in Fig.\ref{fig:FocalSpot}. All ellipse have the same area as mentioned above. The upper ellipse is shaped by microbeam from channel 0. It has greater x-size because the larger distance from the output channel to the focal point. The bottom   ellipse is related to the last microbeam. It has a smallest deformation. The centers of the ellipses in focal plane are located on the $V_x$-axis at the points

\begin{equation}
v_{x,n}=-\frac { v_z sin(\vartheta_n)}{cos(\vartheta_n)}. 
\label{FocusEllipsesCenters}
\end{equation}

For an experiment measurements, it is important to find the intensity profile at the focal spot. It can be received by the integrating Eq.(\ref{TotalDistribution}) over $v_x$: 

\begin{equation}
f_{out}(x,F_{max})=\int\int\sum_{n=0}^{N-1} f_{out,n}(x,F_{max},v_x,v_z) dv_x dv_z. 
\label{TotalFocusProfile}
\end{equation}

In practical calculations, the summation over a pile of channels (see Fig.\ref{fig:FocalSpot})can be change by integrating over $n$.To simplify the calculations, we assume that all $\vartheta_n << 1$. For normalization (\ref{NormalizationDistibutionOneCh}) this gives the pick value of distribution function in the form

\begin{eqnarray}
f_{out}(0,F_{max}) =  
2\,{\it r_{ch}}\,{\it v_z}\, \left( {\frac {\sqrt {{{\it v_z}}^{2}+{{\it D}
}^{2}  \cos \left( \Gamma \right)^{2}}}{{\it D}\,{
\it d_p}\,\sin \left( \Gamma \right) }}\right. -\nonumber\\  
\left.- {\frac {{{\it v_z}}^{2}+{{\it D
}}^{2}  \cos \left( \Gamma \right)^{2}}{{\it D}\,{
\it d_p}\,\sin \left( \Gamma \right) \sqrt {{{\it v_z}}^{2}+{{\it D}}^{
2} \cos \left( \Gamma \right)^{2}-2\,{\it D}\,{\it 
d_p}\,\sin \left( \Gamma \right) N}}} \right) + \nonumber\\
+4\,{\frac {{\it r_{ch}}\,{
\it v_z}\,N}{\sqrt {{{\it v_z}}^{2}+{{\it D}}^{2} \cos \left( 
\Gamma \right)^{2}-2\,{\it D}\,{\it d_p}\,\sin \left( \Gamma
 \right) N}}}, 
\label{IntensityFocus}
\end{eqnarray}

where the following parameters were introduced

\begin{eqnarray}
sin(\Gamma)=\frac {R_{min}}{D}, \nonumber\\
cos(\Gamma)=\frac {\sqrt{D^2-R_{min}^2}}{D}.
\label{ParamsforDistribution}
\end{eqnarray}

Note that in order to calculate the envelope of the beam in phase space, we may take a derivative of microbeam surface with respect to $n$ considering variable $n$ as continuous and  eliminate $n$ from two equations

\begin{eqnarray}
\Phi_{n}(x,z,v_x,v_z) = 0, \nonumber\\
\frac{\partial \Phi_{n}(x,z,v_x,v_z)}{\partial n}  = 0,
\label{Envelope_surface}
\end{eqnarray}

where $\Phi_{n}(x,z,v_x,v_z) =0$ is the equation of the microbeam surface in phase space. We can substitute $z$ for $F_{max}$ (see Eq.(\ref{focal_distances})) in equations (\ref{Envelope_surface}), so getting 

\begin{eqnarray}
\Phi_{n}(x,F_{max},v_x,v_z) = 0, \nonumber\\
\frac{\partial \Phi_{n}(x,F_{max},v_x,v_z)}{\partial n}  = 0,
\label{Envelope_spot}
\end{eqnarray}

the equations of the cross section envelope line at the focal spot.  
If the distribution (\ref{Distribution_n_Chan}) or (\ref{Distribution_All_Chan}) is used as a model, we have the implicit equation of n-th microbeam in focal plane $(x,v_x)$ which actually used to depict Fig.\ref{fig:FocalSpot}: 

\begin{eqnarray}
\Phi_{n}(x,F_{max},v_x,v_z)={\frac { \left( B {\it v_s}- C x \right) ^{2}}{ \left( {\it v_s}\,{\it 
sin(\vartheta_n)}-C \right) ^{2}}}+{{\it v_s}}^{2}-\,{\frac {{{\it r_{ch}}}^{2}}{1-
{{\it sin(\vartheta_n)}}^{2}}}=0.  
\label{Model Surface}
\end{eqnarray}

The coefficients $B, C$ here are given functions of the $\vartheta_n$:

\begin{eqnarray}
C=\frac { v_z}{\sqrt {1-sin(\vartheta_n)^{2}}}, \nonumber\\
B=D \, cos(\Gamma+\vartheta_n), \nonumber\\
v_s= {\it v_x}+  C \,sin(\vartheta_n)
\label{ParamsforModelSurface}
\end{eqnarray}

The calculation of the envelope by Eq.(\ref{Envelope_spot}) with surface (\ref{Model Surface}) are straightforward but cumbersome. In the case $v_s  sin(\vartheta_n)<< C $, which is always true for fast particles,  the envelope has a form of two deformed hyperbolas with asymmetry. The parametric presentation for these curves $(x_1,v_{x,1}), (x_2,v_{x,2})$ are (the parameter for curves is $\vartheta_n $) 

\begin{eqnarray}
\left\{ \begin{array}{ll}
x_1= {\frac {{\it r_{ch}}\,\sqrt {B^2+C^2}}{C cos(\vartheta_n)}}\\
v_{x,1}={\frac {{\it r_{ch}}\,B}{\sqrt {B^2+C^2}cos(\vartheta_n)}}-C{\it sin(\vartheta_n)}\nonumber\\
\end{array} \right. ,\\
\left\{ \begin{array}{ll}
x_2= -{\frac {{\it r_{ch}}\,\sqrt {B^2+C^2}}{C cos(\vartheta_n)}}\\
v_{x,2}=-{\frac {{\it r_{ch}}\,B}{\sqrt {B^2+C^2}cos(\vartheta_n)}}-C{\it sin(\vartheta_n)}
\end{array} \right.
,
\label{EnvelopeParametricPresentation}
\end{eqnarray}

where $B =B(\vartheta_n), C=C(\vartheta_n)$ are defined in Eqs.(\ref{ParamsforModelSurface}).
These formulas are accurate for foci located far away from the crystal as well as for crystalline geometry provided the maximum magnification and minimum focusing size proposed in \cite{kov03}. The envelope calculated from them is shown in Fig. \ref{fig:FocalSpot}.   

\section{\label{sec:level6}Conclusion}

In conclusion, we describe some possible limitations of focusing which can substantially reduce the focusing effect. The smoothness of the cylindrical surface is of great importance, the amorphous layer on that surface must be as thin as
possible, and  all adjacent channels must constitute the staircase
with step widths equals to the interplaner space. Of course, it is not always possible and deviation caused by surface roughness or mosaic blocks with widths about the length of channeling oscillation will produce a broadening of the
focal spot and greatly decrease the peak intensity. Other well known source of
aberration to be considered for focusing is negative Gaussian
curvature of bent crystals (anticlastic effect). The asymmetry of
centripetal dechanneling in bent channels \cite{bck,tar98} is a strong source 
of asymmetry in focal spot. There is also a possibility of asymmetry in
coherent transitional scattering when the particles come out with
small angles and move some short period of time in semichannel. For long crystal, dechanneling plays an important role, and the maximum possible focal distance \cite{kov03} when the focal spot lies almost on the tip of the crystal is limited by this process.

\bibliographystyle{elsart-num}
\bibliography{../../Focusing_and_Channeling_in_Crystals/chan02}% Produces the bibliography via BibTeX.

\end{document}